\documentclass[conference]{IEEEtran}
\IEEEoverridecommandlockouts
\usepackage{cite}
\usepackage{amsmath,amssymb,amsfonts}
\usepackage{algorithmic}
\usepackage{graphicx}
\usepackage{textcomp}
\usepackage{xcolor}
\usepackage{subcaption}
\usepackage{mathtools}
\usepackage{booktabs}
\usepackage{multirow}
\usepackage{adjustbox}
\usepackage{makecell}
\usepackage{booktabs, makecell, xcolor}
\usepackage{mathrsfs}
\usepackage{hyperref}

\def\BibTeX{{\rm B\kern-.05em{\sc i\kern-.025em b}\kern-.08em
    T\kern-.1667em\lower.7ex\hbox{E}\kern-.125emX}}
\begin{document}

\title{SCALED : Surrogate-gradient for Codec-Aware Learning of Downsampling in ABR Streaming}

\author{
\IEEEauthorblockN{
Esteban Pesnel\IEEEauthorrefmark{1}\IEEEauthorrefmark{2}, 
Julien Le Tanou\IEEEauthorrefmark{1}, 
Michael Ropert\IEEEauthorrefmark{1},
Thomas Maugey\IEEEauthorrefmark{2},
Aline Roumy\IEEEauthorrefmark{2}
}
\IEEEauthorblockA{
\IEEEauthorrefmark{1}MediaKind, Rennes, France \\
\{esteban.pesnel, julien.letanou, michael.ropert\}@mediakind.com
}
\IEEEauthorblockA{
\IEEEauthorrefmark{2}INRIA, Rennes, France \\
\{esteban.pesnel, aline.roumy, thomas.maugey\}@inria.fr
}
}

\maketitle

\begin{abstract}
The rapid growth in video consumption has introduced significant challenges to modern streaming architectures. Over-the-Top (OTT) video delivery now predominantly relies on Adaptive Bitrate (ABR) streaming, which dynamically adjusts bitrate and resolution based on client-side constraints such as display capabilities and network bandwidth. This pipeline typically involves downsampling the original high-resolution content, encoding and transmitting it, followed by decoding and upsampling on the client side. Traditionally, these processing stages have been optimized in isolation, leading to suboptimal end-to-end rate-distortion (R-D) performance.
The advent of deep learning has spurred interest in jointly optimizing the ABR pipeline using learned resampling methods. However, training such systems end-to-end remains challenging due to the non-differentiable nature of standard video codecs, which obstructs gradient-based optimization. Recent works have addressed this issue using differentiable proxy models, based either on deep neural networks or hybrid coding schemes with differentiable components such as soft quantization, to approximate the codec behavior. While differentiable proxy codecs have enabled progress in compression-aware learning, they remain approximations that may not fully capture the behavior of standard, non-differentiable codecs. To our knowledge, there is no prior evidence demonstrating the inefficiencies of using standard codecs during training. In this work, we introduce a novel framework that enables end-to-end training with real, non-differentiable codecs by leveraging data-driven surrogate gradients derived from actual compression errors. It facilitates the alignment between training objectives and deployment performance. Experimental results show a 5.19\% improvement in BD-BR (PSNR) compared to codec-agnostic training approaches, consistently across the entire rate-distortion convex hull spanning multiple downsampling ratios.
\end{abstract}

\section{Introduction}
Video consumption has grown exponentially during the last decades, representing more than 70\% of network traffic~\cite{Ericsson}. With the emergence of HTTP-based video streaming solutions such as DASH~\cite{DASH} or HLS~\cite{HLS}, modern video platforms heavily rely on Adaptive Bitrate streaming (ABR), which dynamically adapts the bit rate and resolution of the delivered content to meet client bandwidth requirements and display. The high-resolution content is first downsampled through various scale ratios $s \in \mathbb{Q}^{+}$ and separately encoded. Then, the bit stream is decoded by the client and upsampled to the device display resolution.

Traditional downsampling methods such as bicubic, lanczos or bilinear filters~\cite{bicubic,lanczos} rely on linear content-invariant filters, targeting low-pass filtering prior to subsampling. These filters enjoy low computational complexity, despite visible artifacts (e.g. blurring, ringing, etc.).
Recognizing the limitations of content-invariant designs, non-linear methods adapt to local image structure. For example, bilateral filtering~\cite{bilateral} combines spatial proximity and intensity similarity to perform edge-aware smoothing, preserving fine structures while reducing noise. More recent adaptive downsampling strategies leverage semantic content and visual saliency to guide resolution reduction~\cite{CAID, Perceptual, Rapid-detail, L0}.
While classical methods remain integral to video streaming systems, they treat downsampling independently of subsequent processing stages—particularly upsampling and compression. This decoupling often results in sub-optimal reconstructions, especially when aggressive compression introduces quantization and blocking artifacts. An emerging direction addresses this by co-designing downsamplers with knowledge of the upscaler and codec.

\begin{figure*}[ht]
  \centering
  \includegraphics[width=0.9\textwidth]{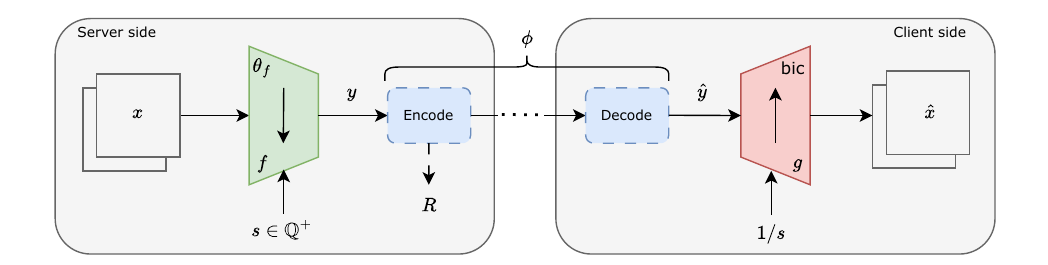}
  \caption{ABR streaming pipeline. The server downsamples and encodes the high-resolution video at scale ratio $s$, then the client decodes and upsamples it to match the display. Blue dashed boxes are non-differentiable; the red box is fixed.}
  \label{fig:wide-image}
  \vspace{-3mm} % <--- Ajustez cette valeur (ex: -0.5cm, -10pt)
\end{figure*}

With the emergence of deep learning, several studies have focused on integrating deep learned pre- or post-processing around traditional codecs. Among these solutions, several works introduced neural network model for resampling tasks. Joint learning of downsampling and upsampling~\cite{CAR} leads to significant improvements over traditional filtering baselines. However, learned upsamplers increase the client-side computational complexity and may conflict with standardization requirements for technology deployment and mass adoption. To mitigate this problem, other studies~\cite{CNN-CR,Chen2024} propose CNN-based learned downsampling based on bicubic upsampler at training, subject to end-to-end MSE loss function minimization. However, these models are trained without considering the compression process due to its non-differentiable nature, which is suboptimal for video streaming applications. Indeed, the non-differentiability poses a challenge for gradient-based optimizers, in order to minimize a convex objective function.

As a result, recent studies have attempted to overcome this issue by modeling the codec with a differentiable proxy. Son \textit{et al.}~\cite{ACN} propose to mimic traditional codec behavior using deep neural networks. More recently, Guleryuz \textit{et al.}~\cite{Google2024} introduced a differentiable codec proxy that implements the classical paradigm of hybrid video coding. Specifically, it incorporates a straight-through quantizer, a motion estimation module derived from UFlow~\cite{UFlow}, and a loop filter proxy based on U-Net~\cite{U-Net}. The usage of a differentiable codec during the training tends to improve the learning of pre- and post-processing networks, such as downsampling and upsampling. However, the training efficiency deeply relies on the codec proxy accuracy. Indeed, learned codec proxies are trained to minimize the MSE with real codecs, which may be insufficient to fully capture the specific characteristics of video codecs. 

In this context, we propose an alternative learning strategy that integrates the ground-truth standard codec into the downsampling optimization. To address the codec's non-differentiability, we introduce a surrogate gradient in the backward pass, defined as the derivative of the standard deviation of the ground-truth compression error, modulated by the compression error itself. While during the forward pass, the actual compression error is directly added into the optimization process. We then propose and assess two learning scheme variants based on this strategy against codec-agnostic training schemes. In the first variant, we aim to minimize the reconstruction distortion by leveraging the actual coding artifacts at a given bit rate. In the second variant, we incorporate a rate constraint into the training loss, leveraging on a differentiable rate proxy from the state of the art to enable efficient rate-distortion optimization. Our contribution outperforms codec agnostic baseline for deep learned downsampling by 5.19\% in BD-BR(PSNR) over then entire bit rate-quality convex hull of different scaling ratios, demonstrating the effectiveness of using standard, non-differentiable codecs during training.

\section{Training scheme for downsampling optimization}

\subsection{Problem formulation}

A typical ABR streaming scheme is illustrated in Figure~\ref{fig:wide-image}. The client first requests a video chunk at a specific bit rate from the server. The high-resolution content $\boldsymbol{x}$ is then downsampled using a scale ratio $s$ and encoded with a standard codec $\phi$. The resulting representation is transmitted to the client, which decodes and upsamples the content to match the display resolution, yielding a reconstructed content $\boldsymbol{\hat{x}}$.

To optimize this video delivery pipeline, we introduce a learned downsampling function with associated weights $f(\boldsymbol{x};\boldsymbol{\theta}_f)$, and adopt a fixed bicubic upsampler $g$ for reconstruction. The selection of a linear fixed filter for the upsampling task is motivated by the will to avoid non-standard and computationally intensive technology at the end-client, enabling broad solution adoption. Learned downsampler $f$ aims at minimizing the end-to-end distortion $D$, for the smallest bit rate budget $R$, given by a ground-truth rate function $\mathrm{r}$.

\vspace{-0.7em} 

\begin{equation}
\boldsymbol{\theta}_{f}^{*} = \arg\min_{\boldsymbol{\theta}_f} \|\boldsymbol{x}-g(\phi(f(\boldsymbol{x};\boldsymbol{\theta}_f)))\|^2_2 + \lambda \mathrm{r}(\phi(f(\boldsymbol{x};\boldsymbol{\theta}_f)))
\label{eq}
\end{equation}

%\begin{equation}
 %   R = \mathrm{r}(\phi(f(\x;\theta_f)))
%\label{eq}
%\end{equation}

However, both $\phi$ and $\mathrm{r}$  are non differentiable, preventing the direct optimization of (1).

\subsection{Existing methods and limitations}

Few studies relaxed this constraint by optimizing the downsampler independently of the codec, which was excluded from the training process. Li \textit{et al.}~\cite{CNN-CR} proposed a 10-layer convolutional neural network (CNN) for image downsampling task. Building on a similar architecture, Chen \textit{et al.}~\cite{Chen2024} extended the approach to support arbitrary fractional downscaling ratios, with $s \in \mathbb{Q}^+$. The original content $\boldsymbol{x}$ is first downsampled. The low-resolution content $\boldsymbol{\tilde{x}}$ is then bicubic upsampled through a linear function $g$. The reconstructed signal is then compared to the original one, leading to a codec-agnostic D-only minimization: 
\vspace{-0.8em} 

\begin{equation}
\boldsymbol{\theta}_{f}^{*} = \arg\min_{\boldsymbol{\theta}_f} \|\boldsymbol{x}-g(f(\boldsymbol{x};\boldsymbol{\theta}_f))\|^2_2
 \label{eq}
\end{equation}

However, training without considering the compression process is sub-optimal for video streaming applications, as both the distortion induced by compression and output bit rate are neglected. 

Hence, codec-aware training scheme becomes essential to obtain significant gain in rate distortion. Thus, some recent works~\cite{ACN,CNN-RD,ByteDance,Shangai}  have attempted to model the codec process in a differentiable manner, by designing $\hat{\phi}$ and $\hat{r}$ proxy functions. A common approach among these methods relies on CNN~\cite{CNN-RD} or U-Net~\cite{ACN} to mimic classical codec behavior. Alternatively, other works introduced codec proxy based on hybrid image or video compression scheme. Xing \textit{et al.}~\cite{AIDN} introduced a JPEG estimator based on DCT and soft quantization. Guleryuz \textit{et al.}~\cite{Google2024} extend this paradigm to video compression by incorporating a motion estimation proxy based on U-Flow~\cite{UFlow}, a loop filter proxy based on U-Net~\cite{U-Net} and a straight-through estimated quantizer~\cite{STE}. 

All of these methods implement differentiable entropy estimators or rate proxies, $\hat{r}$, to enable end-to-end Rate Distortion optimization. Although numerous approaches rely on deep learning models to estimate bit rate~\cite{ACN,ByteDance,Shangai}, some studies exploit the DCT coefficients of the compressed image, modeling the rate through affine mappings~\cite{Google2024}. The corresponding optimization function is :
\vspace{-0.3em} 

\begin{equation}
\boldsymbol{\theta}_{f}^{*} = \arg\min_{\boldsymbol{\theta}_f} \|\boldsymbol{x}-g(\hat{\phi}(f(\boldsymbol{x};\boldsymbol{\theta}_f)))\|^2_2 + \lambda \hat{R}
 \label{eq}
\end{equation}

Here, the set of frames $\boldsymbol{x}$ is first downsampled through $f$. The codec proxy $\hat{\phi}$ then produces a sequence of frames that mimic those of a compressed video. The rate proxy $\hat{r}$ estimates the associated bit rate $\hat{R}$. The result is then bicubic upsampled and compared to $\boldsymbol{x}$. We then obtain a rate-distortion optimization problem, based on codec and rate proxies.

However, another challenge arises from codec proxy training. Most proxy models are optimized using a mean squared error loss.
\vspace{-1em} 

\begin{equation}
\boldsymbol{\theta}_{\hat{\phi}}^{*} = \arg\min_{\theta_{\hat{\phi}}} \|\phi(\boldsymbol{x})-\hat{\phi}(\boldsymbol{x})\|^2_2 
 \label{eq}
\end{equation}

However, the MSE metric may be insufficient to capture the true characteristics and artifacts introduced by the codec, such as ringing or blocking effects. As a result, even if the proxy minimizes the MSE, it may fail to reproduce these codec-specific artifacts accurately. This mismatch is sub-optimal for the optimization of the overall objective (4).

\begin{figure}[t]
  \centering
  \includegraphics[width=0.45\textwidth]{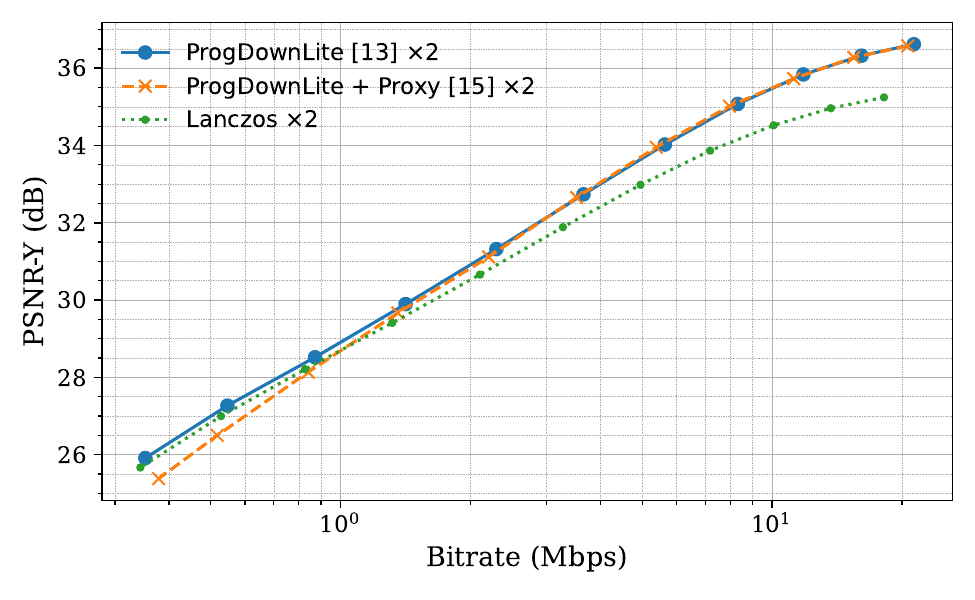}
  \vspace{-2mm} % <--- Ajustez cette valeur (ex: -0.5cm, -10pt)
    \caption{Rate-distortion curves of WestWindEasy sequence with scale $s=1/2$, comparing ProgDownLite~\cite{Chen2024} with and without the codec proxy~\cite{Google2024}, against a Lanczos baseline.}
  \label{RD-curves-1-2}
  \vspace{-5mm} % <--- Ajustez cette valeur (ex: -0.5cm, -10pt)
\end{figure}

More specifically, we tested the codec proxy from sandwiched video compression~\cite{Google2024} in the context of ABR streaming. We trained a deep downsampler alongside the codec proxy, with fixed upsampler for each considered bit rate. %However, we were not able to obtain significant gains in our specific context. 
An example of the obtained rate-distortion curve is shown in Figure~\ref{RD-curves-1-2}. %Once having acceptable results 
We can see that the solution based on codec proxy slightly improves the codec-agnostic solution in high and medium bit rates but obtains lower performance at low bit rate. %, the downsampler trained with the codec proxy were not able to maintain these performances in low bitrates. We made the assumption that the straight-through quantizer is not optimal for low bitrate training. 
%Indeed, since the derivative of quantization process is set to 0, the downsampler weights are not updated according to the degradation introduced by the quantization. However, at low bitrates, quantization has a significant role in the end-to-end distortion. Thus, ignoring the quantization noise during the backward pass may lead to sub-optimal solution. 
This weakness may be explained by several limitations in the proxy design. First, quantization error is neglected during backpropagation. Moreover, prediction proxies are estimated on source, and syntax costs are omitted - making the estimators invariant to the target rate.

Instead of relying on a codec proxy, we opted to directly use a real, non-differentiable codec during training. While codec proxies can facilitate gradient-based optimization, their design is highly non-trivial and must be carefully engineered to accurately emulate the behavior of real encoders. In contrast, real codecs already implement the full set of compression tools and can be readily integrated into the pipeline. This choice enables more faithful supervision and simplifies the training setup, while opening the door to further optimization strategies without requiring additional proxy design efforts. 

%The design constraints of the proxy reduce its accuracy at low bitrates, as they prevent the consideration of critical rate-dependent factors such as syntax overhead or rate-dependant intra/inter prediction decisions.

% Then, learning a deep downsampler $f$ with accurate compression information is a key challenge. Indeed a codec proxy may diverge from the ground truth behavior of the real codec at inference time, resulting in a suboptimal training of the downsampling function $f$.

\section{Proposed approach}

Learning with non-differentiable function is a fundamental research topic in machine learning. Several works have attempted to circumvent this issue, notably using the Straight-Through Estimator (STE)~\cite{STE}, the injection of additive noise 
from uniform distribution~\cite{Balle-ICLR-2017}, or soft-to-hard annealing~\cite{Agustsson-2020}. They all aimed at addressing the non-differentiability introduced by components such as quantization in codecs.
A common way to implement STE, and handle non-differentiable components, is to use the stop-gradient ($\operatorname{sg}$) operator, which explicitly blocks the gradient flow during backpropagation through selected terms of the computation graph. The function $\operatorname{sg}(\boldsymbol{x})$ acts as the identity function in the forward pass but returns a zero gradient in the backward pass.

\begin{equation}
\operatorname{sg}(\boldsymbol{x}) = 
\begin{cases}
\mathcal{F} : \boldsymbol{x} \\
\nabla \mathcal{L} : \dfrac{\partial\,\operatorname{sg}(\boldsymbol{x})}{\partial \boldsymbol{x}} = 0
\end{cases}
\label{eq:stopgrad}
\end{equation}

Here, $\mathcal{F}$ denotes the forward pass and $\nabla\mathcal{L}$ the backward pass. Thus, $\boldsymbol{x}$ is considered as a static constant during the forward pass. 
The application of this operator for STE implementation in the context of codec-aware design is used to detach the compression error $\boldsymbol{\epsilon}$ from the optimization path:

\begin{equation} \boldsymbol{\hat{y}} = \underbrace{f(\boldsymbol{x};\boldsymbol{\theta}_f)}_{\boldsymbol{y}} + \operatorname{sg}(\underbrace{\phi(f(\boldsymbol{x};\boldsymbol{\theta}_f))-f(\boldsymbol{x};\boldsymbol{\theta}_f)}_{\boldsymbol{\epsilon}}) \end{equation}

Here, $\phi$ denotes the true codec function, and $\boldsymbol{\epsilon}$ is treated as additive noise during the forward pass, whose unknown gradient is set to zero during the backward pass.
We explored the use of the Straight-Through Estimator (STE) in our context. However, this approach led to the same challenges identified by Mack \textit{et al.}~\cite{Mack}, which confirms the limitations of regular STE for compression applications.
Indeed, since $\frac{\partial \boldsymbol{\hat{y}}}{\partial \boldsymbol{y}} = 1$, the optimization disconnects the compression error in the gradient flow which leads to an increase in $|f(\boldsymbol{x};\boldsymbol{\theta}_f)|_1$, that prevents from proper convergence during training~\cite{Mack}.

To address this issue, we propose a method based on a modification of the STE from Mack \textit{et al.}~\cite{Mack} to enhance the surrogate gradient of the compression error during back-propagation, as shown in (7) and (8). 

\begin{equation}
    \boldsymbol{\hat{y}} = \boldsymbol{y} + \operatorname{sg}(\boldsymbol{\epsilon}) \left[ \frac{\sigma (\boldsymbol{\epsilon})}{\operatorname{sg}(\sigma(\boldsymbol{\epsilon}))} \right]
\end{equation}

 We denote by $\sigma$ the standard deviation operator, and by $\boldsymbol{\hat{y}}$ the encoded version of the downsampled content with the proposed equation. The modified-STE corresponds to a reparameterization  trick~\cite{Kigma} to account for the compression noise in the backpropagation gradient, resulting in the definition of an enhanced surrogate gradient based on the derivative of the standard deviation of the true compression error, modulated by the compression error itself.

\vspace{-0.2cm}
\begin{equation}
     \left( \frac{\partial\boldsymbol{\hat{y}}}{\partial \boldsymbol{y}} \right)_{\text{surrogate}}
     = \mathbf{I} + \frac{\boldsymbol{\epsilon}}{\sigma(\boldsymbol{\epsilon})}\frac{\partial{\sigma(\boldsymbol{\epsilon})}}{\partial \boldsymbol{y}} = \mathbf{I} - \frac{\boldsymbol{\epsilon} (\boldsymbol{\epsilon} - \boldsymbol{\bar{\epsilon}})^T}{N\sigma^2(\boldsymbol{\epsilon})}
\label{eq:stat_jacobian}
\end{equation}

%, as shown below with a matrix formulation.
%\begin{equation}
%    \frac{\partial\sigma(\epsilon)}{\partial f} = -\frac{\epsilon - %\bar{\epsilon}}{n \cdot \sigma(\epsilon)}
%\end{equation}
% Here $P$ is a transition matrix and $n$ the number of samples. 
% As a result, the surrogate gradient of the compression error is set to: 

% \begin{equation}
%     \frac{\partial \boldsymbol{\epsilon}}{\partial f} \approx \frac{\boldsymbol{\epsilon}}{\sigma(\boldsymbol{\epsilon})}\odot \frac{\partial \sigma(\boldsymbol{\epsilon})}{\partial f}
% \end{equation}
where $\boldsymbol{\bar{\epsilon}}$ is a vector containing the average value of $\boldsymbol{\epsilon}$.

Thus, we now approximate, during backpropagation, how changes in input $\boldsymbol{y}$ influence the compression error, using its magnitude and statistical spread.
The term $\frac{\boldsymbol{\epsilon}}{\sigma(\boldsymbol{\epsilon})}$ acts as a normalized measure of the quantization noise, that stabilizes gradient flow, while the derivative of sigma w.r.t. $\boldsymbol{y}$ reflects how the noise distribution changes with $\boldsymbol{y}$. This gives a dynamically scaled surrogate gradient for $\boldsymbol{\epsilon}$ sensitive to compression error variability. Then, leveraging on (7), we will consider the two following proposals with true compression errors: 
\begin{itemize}
    \item an MSE-based training scheme with non-differentiable codec, based on (2) and (7).

\begin{table*}[h]
\centering
\renewcommand{\arraystretch}{1.2}
\scalebox{0.92}{
\begin{tabular}{cccccccccccc}
\toprule
\textbf{Filter} & \makecell{\textbf{Training}\\\textbf{dataset}} &  \makecell{\textbf{Training}\\\textbf{strategy}} & \makecell{\textbf{Codec}\\\textbf{aware}} 
& \multicolumn{4}{c}{\textbf{XIPH}} 
& \multicolumn{4}{c}{\textbf{UVG}~\cite{UVG}} \\
\cmidrule(lr){5-8} \cmidrule(lr){9-12}
& & & & \textbf{PSNR} & \textbf{SSIM} & \textbf{VMAF} & \textbf{VMAF\textsubscript{NEG}} 
& \textbf{PSNR} & \textbf{SSIM} & \textbf{VMAF} & \textbf{VMAF\textsubscript{NEG}} \\
\midrule
ProgDownLite\textsubscript{RGB} & DIV2K & D-only (2) & No & -0.66 & -1.24 & -5.67 & -3.58 & 1.22 & -1.78 & -7.72 & -5.24 \\
ProgDownLite\textsubscript{YUV} & Google~\cite{Google2024} & D-only (2) & No & -0.85 & -1.19 & -4.71 & -3.60 & 1.91 & -2.31 & -6.11 & -5.02 \\
ProgDownLite\textsubscript{YUV} & Google~\cite{Google2024} & STE (6) & Yes & 68.57 & 48.34 & 11.22 & 20.69 & 49.39 & 45.33 & 4.46 & 5.06 \\
ProgDownLite\textsubscript{YUV} & Google~\cite{Google2024} & Proxy \cite{Google2024} & Yes
& -1.47 & -1.87 & -2.58 & -1.67
& -0.08 & -2.99 & -5.88 & -4.79 \\
ProgDownLite\textsubscript{YUV} & Google~\cite{Google2024} & SCALED\textsubscript{D} (10) & Yes & -4.67 & \textbf{-5.21} & \textbf{-9.80} & \textbf{-7.78} & -2.84 & \textbf{-5.79} & \textbf{-10.95} & \textbf{-9.07} \\
ProgDownLite\textsubscript{YUV} & Google~\cite{Google2024} & SCALED\textsubscript{RD} (11) & Yes & \textbf{-6.11} & -4.29 & -8.80 & -7.59 & \textbf{-5.07} & -5.21 & -8.88 & -8.02 \\
\bottomrule
\end{tabular}
}
\caption{Comparison of filtering methods on the XIPH and UVG datasets. Metrics are computed as BD-BR values with respect to the convex hull of the Lanczos filter, used as the reference rate-distortion baseline.}
\label{tab:filters_xiph_uvg}
\end{table*}
    \item a Rate-Distortion training scheme with non-differentiable codec, based on (3) and (7).
\end{itemize}

\subsection{Distortion based model with non-differentiable codec}\label{SCM}

In this variant, (7) is substituted in (2), resulting in the new optimization formulation presented in (10):
\vspace{-1em} 

\begin{equation}
\boldsymbol{\theta}_{f}^{*} = \arg\min_{\boldsymbol{\theta}_f} \|\boldsymbol{x}-g(f(\boldsymbol{x};\boldsymbol{\theta}_f)+ \operatorname{sg}(\boldsymbol{\epsilon})\cdot\frac{\sigma(\boldsymbol{\epsilon})}{\operatorname{sg}(\sigma(\boldsymbol{\epsilon}))})\|^2_2
\label{eq}
\end{equation}

Where $\boldsymbol{\epsilon} = \phi(f(\boldsymbol{x};\boldsymbol{\theta}_f))-f(\boldsymbol{x};\boldsymbol{\theta}_f)$, $\phi$ denotes the true compression function. The new network trained using formulation (10) will be referred to as SCALED\textsubscript{D} in the following sections, minimized subject to a certain bit rate budget given to the true encoder.

\subsection{Rate-distortion based model with non-differentiable codec}\label{SCM}

To account for the trade-off between distortion and rate, we now formulate an optimization objective based on the $D + \lambda \hat{R}$ trade-off. We then leverage on (7), (3) and (1), resulting in the new optimization formulation presented in (11)
\vspace{-1em} 

\begin{equation}
\boldsymbol{\theta}_{f}^{*} = \arg\min_{\boldsymbol{\theta}_f} \|\boldsymbol{x}-g(f(\boldsymbol{x};\boldsymbol{\theta}_f)+ \operatorname{sg}(\boldsymbol{\epsilon})\cdot\frac{\sigma(\boldsymbol{\epsilon})}{\operatorname{sg}(\sigma(\boldsymbol{\epsilon}))})\|^2_2 + \lambda\hat{R}
\label{eq}
\end{equation}

Here, $\hat{R}$ denotes a state-of-the-art rate estimation module from Google Sandwiched Compression~\cite{Google2024}, based on differentiable approximation of the $\ell_0$ norm of DCT coefficients. The objective of this variant is to demonstrate the benefits of rate-distortion optimization. The new network trained using formulation (11) will be referred to as SCALED\textsubscript{RD} in the following sections.

\section{Experiments}
\subsection{Implementation details}

To ensure the practical relevance of our learning framework, we base our experiments on a codec that reflects real-world deployment scenarios. We therefore adopt the H.264 standard via its open-source implementation x264~\cite{x264}. We use default medium preset, both for training and experiments.

We adopt the ProgDownLite network~\cite{Chen2024} as a learned downsampling function $f$. This model is based on a five-layer convolutional network for image pre-processing, followed by bilinear downsampling. Subsequently, five additional convolutional layers aim to restore the image details that were suppressed during filtering. Since the resampling is performed by a bilinear downsampling, it supports non-integer scale ratio $s \in \mathbb{Q}^+$, which is mandatory in numerous applications. We fixed the upsampler to a bicubic upsampling.

We use the video dataset from Google%\footnote{via \url{https://github.com/google/sandwiched_compression}}~\cite{Google2024}
, which contains 800 video sequences, in YUV444 colorspace. Each one has a resolution of 256×256 pixels and a length of 10 frames. The chroma channels are first bilinearly downsampled and then bicubically upsampled to simulate the chroma degradation of YUV 4:2:0 content used in the experimental dataset.

\subsection{Performance assessment with compression}

We consider a multi-scale ratio compression scheme, where $s\in \{2/3, 1/2, 2/5, 1/3, 1/4, 1/5\}$. The downsampled contents are then encoded with x264, with the quantization parameter (QP) varying from 17 to 47 in increments of 3. The downsampled, encoded content is then decoded and upsampled back to the original content resolution. We then computed the convex hull of best operating rate-distortion points across the scale ratios set. We trained one SCALED model for each QP and scale ratios.

To assess the downsampling filters performance in rate-distortion context, we select 24 videos from Xiph dataset, and 7 videos from UVG dataset~\cite{UVG}, all in 1920$\times$1080 spatial resolution. For models trained with RGB dataset, each YUV420 video was prior converted to RGB using FFmpeg. 

Table I presents the rate-distortion evaluations. We first reproduce the training conditions of ProgDownLite~\cite{Chen2024} using DIV2K dataset, referred to as ProgDownLite\textsubscript{RGB}. We also re-train ProgDownLite on sandwiched video dataset~\cite{Google2024}, denoted ProgDownLite\textsubscript{YUV}. Finally, we include the results of our two main contributions, SCALED\textsubscript{D} and SCALED\textsubscript{RD}, both trained in the context of non-differentiable codecs and video dataset, and compared to traditional STE. All the models are trained on 100 epochs, with $lr=1e^{-4}$ and $(\beta_1, \beta_2) = (0.9,0.999)$.

It is worth noting that codec agnostic solutions show marginal gains over Lanczos baseline. Indeed, due to the lack of codec consideration at the training, they tend to maintain high frequencies costly to code, resulting in a substantial rate increase. On the other hand, ProgDownLite\textsubscript{YUV} slightly improves the MSE of ProgDownLite\textsubscript{RGB} due to the training dataset nature - videos instead of pictures. 

Then, our SCALED contributions outperform codec-agnostic solutions, thanks to the consideration of the actual codec. Applying a rate constraint with SCALED\textsubscript{RD} increase the performance in MSE, though slighlty degrading other metrics.

\section{Conclusion}

In this paper, we introduce SCALED, a coding-aware downsampling learning framework that leverages data-driven surrogate gradients derived from actual compression errors to optimize the downsampling process end-to-end. The proposed approach demonstrates that surrogate gradients can effectively enable the integration of non-differentiable, real codecs into end-to-end training pipelines. Furthermore, we proposed a hybrid approach in which distortion is estimated using the actual codec, while the rate is approximated via a state-of-the-art differentiable rate proxy. This combination yields improved performance over purely distortion-based methods. Our findings suggest promising directions for developing deep downsamplers, or any learned pre-processing, that are explicitly optimized with respect to real-world codec behavior.

\bibliographystyle{IEEEtran}
\bibliography{refs}

\end{document}